# Polynomial epidemics and clustering in contact networks

Balázs Szendrői (corresponding author)
Lecturer and Researcher
Department of Mathematics, Utrecht University
PO. Box 80010, NL-3508 TA, Utrecht
The Netherlands
Tel: +31 30 253 1186
Fax: +31 30 251 8394
Email: szendroi@math.uu.nl

Gábor Csányi
Research Associate
TCM Group, Cavendish Laboratory, University of Cambridge
Madingley Road, Cambridge CB3 0HE
United Kingdom
Tel: +44 1223 337456
Fax: +44 1223 337356
Email: gabor@csanyi.net

Summary: It is widely known that the spread of the HIV virus was slower than exponential in several populations, even at the very beginning of the epidemic. We show that this implies a significant reduction in the effective reproductive rate of the epidemic, and describe a general mechanism, related to the clustering properties of the disease transmission network, capable of explaining this reduction. Our considerations provide a new angle on polynomial epidemic processes, and may have implications for the choice of strategy against such epidemics.

Our purpose in this note is to discuss polynomial epidemic growth in general terms, drawing attention to a particular aspect which in our view has hitherto been neglected. As a guiding example, consider Figure 1, which plots the growth of the HIV epidemic in different populations. In a number of cases one observes polynomial growth (May & Anderson 1987; Colgate et al 1989): over a long time period, the total number of infected individuals $I(t)$ at time $t$ is well described by a polynomial function $const * t^n$ for some integer number $n$. Such a relationship is valid even at early stages of the epidemic, when global saturation effects could not yet play a role, and when there was no effective intervention, as in US cities in the early 1980s or in Kenya in the late 1980s.

The two fundamental parameters governing disease dynamics (Anderson & May 1991; Hethcote 2000) that are central to our discussion are the basic reproductive rate $R_0$, the number of new infections a host would produce in a totally susceptible population, and the effective reproductive rate $R$ (also called infectee number or replacement number), the average number of actual new infections produced by a host among his contacts during the epidemic process. These quantities satisfy the inequality $R<=R_0$.

The polynomial growth of the HIV/AIDS epidemic cannot be explained by saturation effects (Yorke et al 1978), because it is observed at the earliest stages. The first explanation was proposed by (May & Anderson 1988), arguing that heterogeneity in the distribution of contacts significantly changes the dynamics of a contact epidemic. Models incorporating extreme heterogeneous mixing lead to sub-exponential epidemic curves, as the highly active classes begin to saturate and new infections come from the slower dissemination of infection to less active individuals. In (Colgate et al 1989) it was shown that by making strenuous assumptions about the kind and distribution of the number of contacts, and postulating relatively small and independent subgroups within the population, a polynomial growth rate can indeed be recovered.

Here we propose an alternative point of view. A simple argument shows that for an epidemic that is observed to spread polynomially over a long period of time, the value of the effective reproductive rate $R$ is strongly constrained. Indeed, if $R > 1+c$, where $c$ is a positive constant, the number of infections after time $t$ will exceed $(1+c)^t$ and the disease spreads exponentially. If $R<1-c$, then of course the epidemic withers out. Consequently, if the epidemic is observed to spread slower than exponentially without disappearing altogether, $R$ must approach unity.

We conclude in particular that in all the polynomial epidemic processes of Figure 1, the average number of new HIV infections must approach one per infected host. This surprising and counter-intuitive result is in stark contrast with estimates of (May & Anderson 1987), putting the basic reproductive rate $R_0$ of HIV well above unity. This raises the problem of finding the mechanisms responsible for the reduction from a high value of $R_0$ to the effective $R=1$ required for polynomial growth. How is it possible that in a number if different populations, which presumably have different sexual contact structures, every HIV-positive individual on average infects only one new person?

The transmission of a contact disease like HIV/AIDS is constrained by the network of connections along which transmission is possible (Klovdahl 1985; Potterat et al 1999; Lloyd & May 2001). This opens up the road to use network models to simulate the spread of epidemics (Pastor-Satorras & Vespignani 2001, Lloyd & May 2001). The epidemic curves of Figure 1 set the challenge to find network models giving rise to an effective $R=1$ and subexponential spreading in epidemic processes without recovery such as the HIV epidemic. It is easily shown that regular lattices, as well as random spatially constrained networks with homogeneous spatial distribution, give rise to an effective $R=1$ and polynomial spreading curves. However, such models clearly do not constitute reasonable models of human sexual interactions. We know of no random network construction to date which constitutes a reasonable model of human sexual contacts and which gives polynomial epidemic curves. Note especially that no small-world network, such the preferential attachment scale-free network model of (Barabási & Albert 2001), is suitable: such a network necessarily produces exponential spreading in epidemic processes without recovery.

There is an important network characteristic, not emphasized in earlier work, that plays a role in contact epidemic dynamics: the presence of clustering in the transmission network. We propose that network clustering can have a sufficient local effect on epidemic dynamics which can quench the reproduction rate from a high $R_0$ value to a local effective $R=1$ throughout the population. The models and arguments in (Yorke et al 1987; May & Anderson 1988; Colgate et al 1989, Pastor-Satorras & Vespignani 2001, Lloyd & May 2001) all disregard network clustering.

A network is said to be clustered, if it contains many more short cycles (triangles, or in the case of heterosexual contact networks, cycles of length four) than its other characteristics would imply. The presence of many triangles can equivalently be formulated as the statement that two random contacts of a node are also connected to each other with high probability. One obtains a quantitative measure of clustering in a network by calculating the local clustering coefficient of a node as the ratio of the number of connections between neighbours of the node and the total possible number of neighbour pairs; the clustering coefficient $C$ of the whole network is then obtained as the average of the local clustering coefficients over all the nodes. This measure easily generalizes to cycles of length four in the case of heterosexual networks. Many social and biological networks are known to be clustered, with a value of $C$ several orders of magnitude higher than random networks of equal size and edge density (Watts & Strogatz 1998), and there are strong indications that human social and sexual networks also share this property (Rotherberg et al 1988).

The important point to note is that in a clustered network, the contacts of an infected individual do not form a random sample of the population, in contrast to the usual assumption in epidemic modeling. The proportion of infected persons in this sample is much higher than in the population at large, even early in an epidemic. In other words, the number of non-infected individuals in the immediate neighbourhood of infecting agents is strongly constrained (Keeling et al 1997; Read & Keeling 2003). This leads to a significant reduction in the reproductive rate, possibly capable of turning an $R_0$ which is well over unity into an effective rate $R=1$, required by the observed polynomial growth (see Figure 2).

The proposal that network clustering plays a role in polynomial epidemics is perhaps testable given sufficient data in a well documented disease population. Using the links of the known interaction network and transmission routes, it would be possible to compare the competing effects of contact heterogeneity and clustering directly. It would be of particular interest to study risk networks and epidemic spread in Sub-Saharan Africa, and we hope that relevant data will become available soon.

If our proposal is correct, the observed time development of a contact disease correlates with the clustering properties of its transmission network. The exponential spread (Figure 1) of the HIV epidemic in Eastern Europe is compatible with the fact that its principal transmission mechanism is known to be needle sharing among intravenous drug users (UNAIDS 2000). Indeed, there are indications that needle sharing leads to a network which is significantly less clustered than social or sexual networks (Rothenberg et al 1988). In contrast, the sub-exponential spread of HIV in Sub-Saharan Africa requires an infection mechanism satisfying the constraint that the average number of new infections per host is one. The only mechanism compatible with this is likely to be sexual contact (Walker et al 2003), with a highly clustered transmission network. We predict a high abundance of multiple infections in these populations; therefore, although drug treatment improves the life of HIV-infected individuals, investing in the encouragement of safe-sex practices is likely to be a far better strategy against the epidemic.

**Acknowledgements** The authors wish to thank Bryan Grenfell, Matthew Keeling, John Potterat and especially Robert May for comments and suggestions.

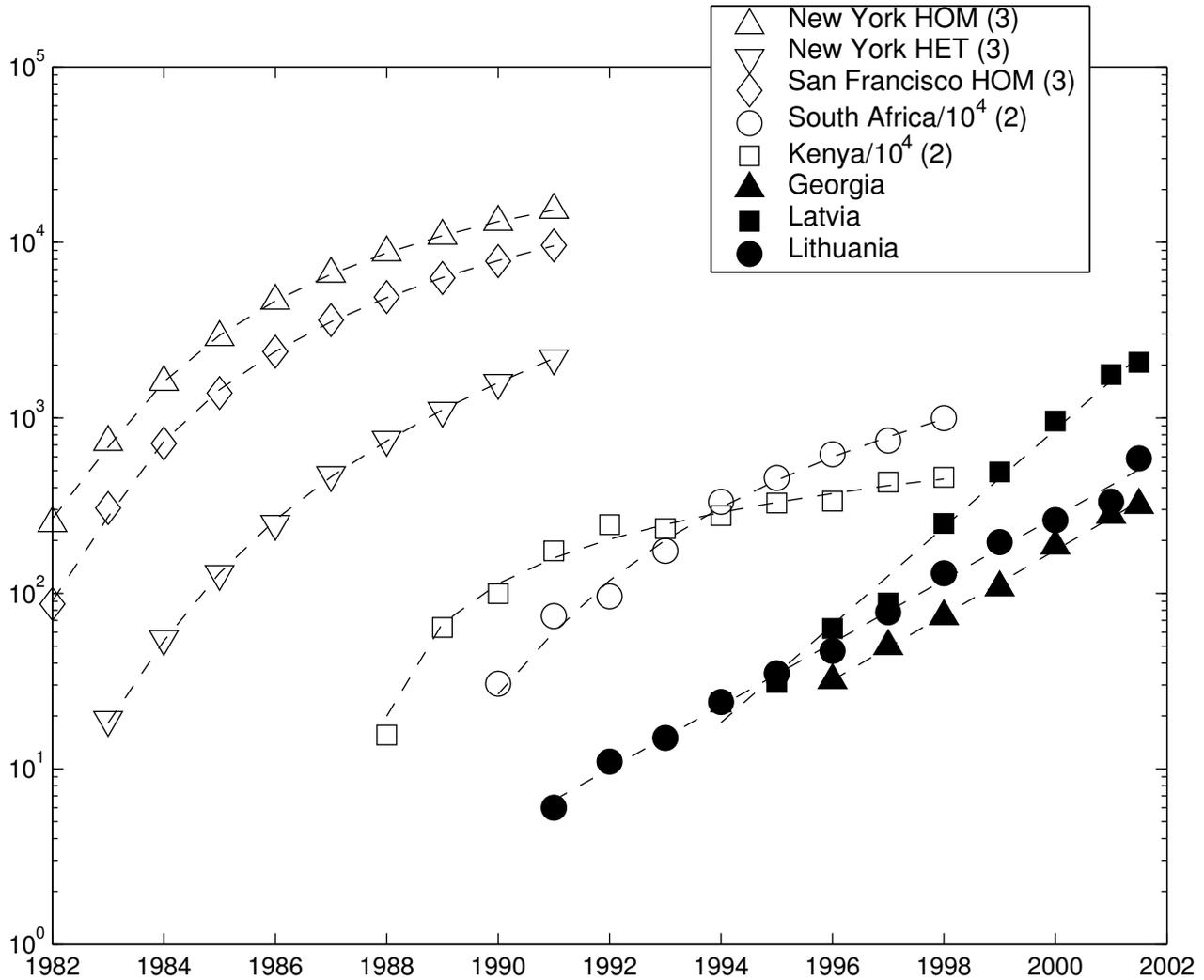

**Figure 1** *The time development of the HIV epidemic on a semi-logarithmic plot in several populations. In US cities (cumulative total of reported HIV infections among homosexuals (HOM) and heterosexuals (HET)) and Sub-Saharan African countries (number of HIV infected individuals, estimated from prevalence among women attending antenatal clinics) we observe a sub-exponential growth, shown by the strong curvature in the plot, well fitted with a polynomial curve (empty symbols, degree of polynomial in parentheses). In Eastern European countries (cumulative total of reported HIV infections), the epidemic is exponential, as shown by the straight lines of the plot (full symbols). (US data from Centers for Disease Control and Prevention, South African data from the Department of Health of the Republic of South Africa, all other data from UNAIDS/EUROHIV databases.)*

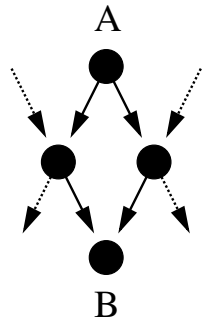

**Figure 2** *Example of the connection between clustering and multiple infections. Arrows indicate the passage of the infection. In this example, A passed on the infection to two individuals, but most people are multiply infected; B gets the infection from two sources. This local phenomenon reduces a global $R_0$ value near 2 to an the effective reproductive rate R close to 1. Note that due to clustering, the contacts of an infected person do not form a random sample of the population, but are much more likely to be infected than the general population at a given time.*

**Short title for page headings**  Polynomial epidemics and clustering